\pdfoutput=1
\documentclass[aps,prx,article,twocolumn,preprintnumbers,amsmath,amssymb,superscriptaddress]{revtex4-2}
\date{\today}
\usepackage{epsfig}
\usepackage{subfigure}
\usepackage{graphicx}
\usepackage{dcolumn}
\usepackage{bm}
\usepackage[colorlinks,citecolor = blue,linkcolor=blue,hyperindex,CJKbookmarks]{hyperref}
\usepackage{float}
\usepackage{hyperref}
\usepackage{comment}
\usepackage{braket}
\newcommand{\aF}    {{\alpha^2F}}
\newcommand{\tildeaF}    {{\alpha^2\Tilde{F}}}

\newcommand{\rvs}[1]{{\color[rgb]{0,0,0}{#1}}}

\begin{document}
\title{Dynamical Approach to Realize Room-Temperature Superconductivity in LaH$_{10}$}
\author{Chendi Xie}\thanks{C.X. and A.D.S. contributed equally to this work.}
\affiliation{Department of Physics and Astronomy, Clemson University, Clemson, SC 29634, USA}
\affiliation{Department of Chemistry, Emory University, Atlanta, GA 30322, USA}
\author{Adam D. Smith}\thanks{C.X. and A.D.S. contributed equally to this work.}
\affiliation{Department of Physics, the University of Alabama at Birmingham, Birmingham, AL 35233, USA}
\author{Haoran Yan}
\affiliation{Department of Chemistry, Emory University, Atlanta, GA 30322, USA}
\author{Wei-Chih Chen}
\affiliation{Department of Physics and Astronomy, Clemson University, Clemson, SC 29634, USA}
\author{Yao Wang}
\email[\href{mailto:yao.wang@emory.edu}{yao.wang@emory.edu}]{}
\affiliation{Department of Chemistry, Emory University, Atlanta, GA 30322, USA}
%\affiliation{Department of Physics and Astronomy, Clemson University, Clemson, SC 29634, USA}

\date{\today}
\begin{abstract}
    Metallic hydrogen and hydride materials stand as promising avenues to achieve room-temperature superconductivity. Characterized by their high phonon frequencies and moderate coupling strengths, several high-pressure hydrides were theoretically predicted to exhibit transition temperatures ($T_c$) exceeding 250\,K, a claim further substantiated by experimental evidence. In an effort to push $T_c$ beyond room temperature, we introduce a dynamical method that involves stimulating hydrides with mid-infrared lasers. Employing Floquet first-principles simulations, we observe that in a nonequilibrium state induced by light, both the electronic density of states and the coupling to high-energy phonons see notable enhancements. These simultaneous improvements collectively result in an estimated 20\%-30\% rise in $T_c$ in practical pump conditions. Our theoretical investigation, therefore, offers a novel strategy to potentially raise the $T_c$ of hydrides above room temperature.
\end{abstract}

\maketitle

\section{Introduction}\label{introduction}

Superconductivity is one of the most important quantum phases discovered to date and is well-known for its promising applications in fields like dissipationless power transmission, quantum computing, and imaging. Due to the strict conditions required for the formation of condensed Cooper pairs, achieving the superconductivity (SC) phase typically requires extremely low temperatures, constrained by the empirical McMillan limit\,\cite{mcmillan1968transition}. To increase the transition temperature $T_c$, previous studies have identified two promising directions. On the one hand, unconventional superconductors, especially cuprates, can exhibit relatively high $T_c$ $\sim$ 130\,K\,\cite{huang1993superconductivity}, while the underlying pairing mechanism and design principle are still under debate. On the other hand, the $T_c$ of conventional superconductors, describable by the Bardeen-Cooper-Schrieffer (BCS) theory, is primarily determined by phonon energies\,\cite{bardeen1957microscopic,bardeen1957theory,esterlis2018a}. This design principle has motivated the idea of leveraging the high-frequency lattice vibrations of the lightest element --- hydrogen\,\cite{ashcroft1968metallic}. Metallic hydrogen has been predicted to have $T_c$ on the order of room temperature but requires extreme pressure over 400\,GPa to stabilize its crystal structure, which has proven experimentally challenging\,\cite{mcmahon2012the,mcminis2015molecular,dias2017observation}. 

Hydrogen-rich compounds, or hydrides, offer a more feasible alternative due to their chemically precompressed lattices\,\cite{ashcroft2004hydrogen,feng2006structures,eremets2008superconductivity}. Various binary hydride compounds have been predicted as candidates for high-temperature superconductors, such as La-H\,\cite{liu2017potential}, Y-H\,\cite{liu2017potential}, H-S\,\cite{li2014the}, Ca-H\,\cite{wang2012superconductive}, Th-H\,\cite{kvashnin2018high}, Ac-H\,\cite{semenok2018actinium} and Hf-H\,\cite{xie2020hydrogen}. To date, high-temperature superconducting signatures have been successfully experimentally verified in H-S\,\cite{drozdov2015conventional}, La-H\,\cite{drozdov2019superconductivity,somayazulu2019evidence,sun2021high}, Y-H\,\cite{kong2021superconductivity} and Ca-H\,\cite{ma2022high}. For example, H$_3$S has been found to have a $T_c$ of 203\,K at 155\,GPa\,\cite{drozdov2015conventional}; more recently, LaH$_{10}$ has exhibited record-breaking $T_c$ of 250\,K at 150\,GPa, 260\,K at 170-200\,GPa and 246\,K at 136\,GPa\,\cite{drozdov2019superconductivity,somayazulu2019evidence,sun2021high}. These findings mark significant strides toward room-temperature SC\,\cite{flores2020a}.

In parallel to the high-pressure synthesis of hydride superconductors, alternative approaches have been considered to enhance $T_c$ of existing superconductors. Among these approaches, ultrafast laser engineering in electronic and phononic structures has emerged as a promising method. Within the pump-probe regime, existing experiments have revealed the possibility of transiently enhancing $T_c$ in K$_3$C$_{60}$\,\cite{mitrano2016possible,cantaluppi2018pressure,budden2021evidence,rowe2023resonant}, which is likely driven by dynamical phonons\,\cite{sentef2016theory,knap2016dynamical,murakami2017nonequilibrium,babadi2017theory,kennes2017transient}. Laser-enhanced SC has also been observed in unconventional superconductors, including La$_{1.675}$Eu$_{0.2}$Sr$_{0.125}$CuO$_4$\,\cite{fausti2011light}, La$_{2-x}$Ba$_x$CuO$_4$\,\cite{nicoletti2014optically}, YBa$_2$Cu$_3$O$_{6.5}$\,\cite{hu2014optically,mankowsky2014nonlinear} and $\kappa$-salts\,\cite{buzzi2020photomolecular,buzzi2021phase}. While these observations may arise from different mechanisms\,\cite{denny2015proposed,raines2015enhancement,patel2016light,nava2018cooling,kennes2019light,michael2020parametric,wang2021fluctuating}, the success in various materials has fueled broad interest in using light to increase the $T_c$ of existing superconductors, and may further extend to hydrides as the final step toward room-temperature SC.

In this paper, we present a dynamical method utilizing mid-infrared (MIR) lasers to raise the transition temperature ($T_c$) of hydrides above 300\,K. Focusing on a light-driven steady state, we simulate the nonequilibrium change in electronic structure and interactions through an integration of Floquet theory and density functional perturbation theory (DFPT), using LaH$_{10}$ as a representative example. Our simulations reveal that the light-engineered electronic structure not only enhances the DOS but also reshapes the Fermi surface, leading to an uneven enhancement electron-phonon coupling (EPC), as depicted by the Eliashberg spectral function. The synergy of these light-engineered properties collectively raises $T_c$. Within a practical range of pump fluences, our simulations \rvs{suggest a possible} a 20\%-30\% increase in $T_c$, exceeding room temperature (300\,K). Additionally, we extend our analysis to evaluate the impact of pressure in LaH$_{10}$ and laser polarization on the light-enhanced $T_c$.

\begin{figure}[!t]
    \centering
    \includegraphics[width=8.2cm]{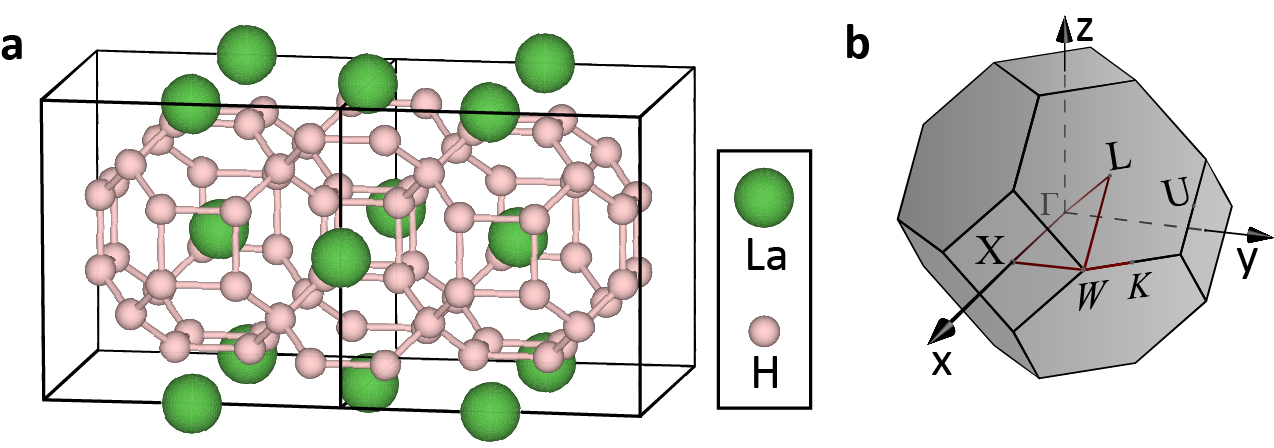	}\vspace{-3mm}
    \caption{\label{fig.wannier}
        \textbf{Structures of fcc LaH$_{10}$.} \textbf{a} Schematic showing the crystal structure of \textit{Fm$\bar{3}$m} (fcc) LaH$_{10}$, showcasing two horizontal unit cells. The La and H atoms are represented by green and red spheres, respectively. \textbf{b} Illustration of the first Brillouin zone in the reciprocal space of fcc LaH$_{10}$, highlighting key high-symmetry momentum points (gray) and momentum cuts (red) referenced in this paper.
    }    
\end{figure}

\begin{figure*}[!t]
    \centering
    \includegraphics[width=1\textwidth]{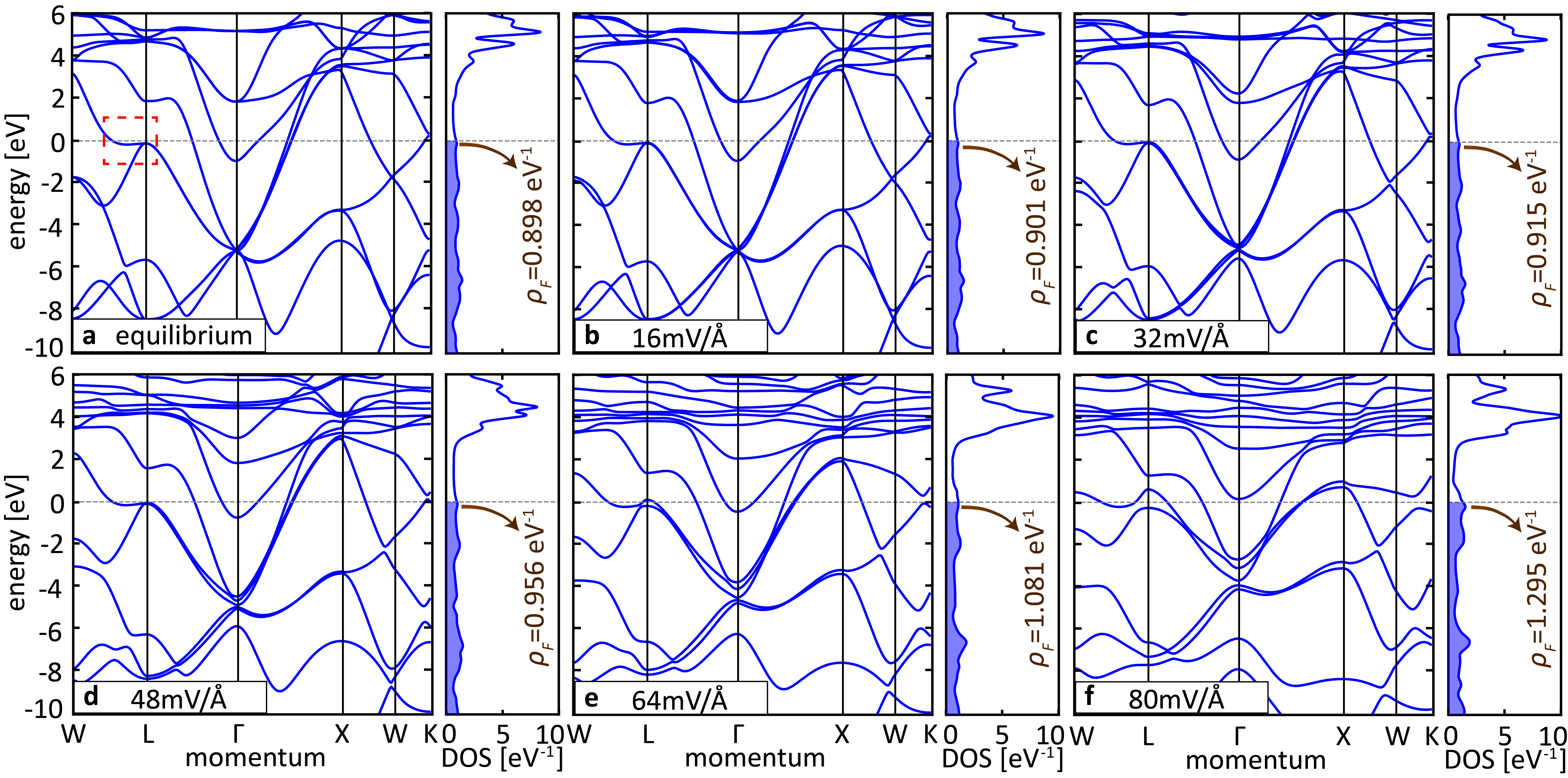	}\vspace{-2mm}
    \caption{\label{fig.band}
        \textbf{Floquet renormalized electronic structures for LaH$_{10}$ under 220\,GPa.} \textbf{a} Wannier-interpolated band structures across the high-symmetry points (left) and electronic density of states (DOS, right) at equilibrium. The DOS at the Fermi level (gray dashed line) $\rho_F$ and the van Hove singularity (vHS) are highlighted, respectively. \textbf{b}-\textbf{f} Same as \textbf{a} but for Floquet steady states induced by an $\hat{x}$-polarized pump, with peak fields $E_0$ set to 16, 32, 48, 64 and 80\,mV/\AA, respectively. 
    }    
\end{figure*}

\section{Results}\label{results}

\subsection{Electronic Structure of Light-Driven LaH$_{10}$}\label{sec.a}

We selected LaH$_{10}$, a BCS-based high-temperature superconductor as the platform for our simulation due to its demonstrated consistency between theoretical predictions\,\cite{liu2017potential,peng2017hydrogen} and experimental findings\,\cite{drozdov2019superconductivity,somayazulu2019evidence,sun2021high}. The crystal structure of LaH$_{10}$ transitions from \textit{R$\bar{3}$m} to \textit{Fm$\bar{3}$m} (fcc) symmetry when pressure increases beyond 150\,GPa\,\cite{liu2017potential,geballe2018synthesis}. As shown in Fig.~\ref{fig.wannier}, its fcc crystal structure displays a highly-symmetric clathrate-like geometry formed by a cage of hydrogen atoms, with individual lanthanum atoms in the cavities. Under 210\,GPa, the first-principles electronic structure and Migdal-Eliashberg theory predict its $T_c$ to be between 264\,K\, and 286\,K\,\cite{liu2017potential,quan2019compressed,peng2017hydrogen}. Soon after these predictions, experiments verified them by observing $T_c = 251$\,K at 168\,GPa\,\cite{drozdov2019superconductivity}, $T_c = 260$\,K at 180\,GPa\,\cite{somayazulu2019evidence} and $T_c = 246$\,K at 136\,GPa\,\cite{sun2021high}. These agreements in structure and transport properties validate the use of first-principles and perturbation simulations for LaH$_{10}$ superconductors. 

To simulate light-matter interactions and dynamical properties, we constructed a tight-binding model using the DFT-simulated Kohn-Sham band structure, with details listed in \textbf{Methods}. Balancing simulation complexity with the clarity of SC description, we constrain the tight-binding model to 25 maximally-localized Wannier orbitals per unit cell and downfold the electronic structure to the second-quantized electronic Hamiltonian
\begin{equation}\label{eq:TBHam}
    \mathcal{H} = \sum_{jl}\sum_{\alpha\beta\sigma} H_{jl}^{(\alpha\beta)}(c_{l\beta\sigma}^\dagger c_{j\alpha\sigma}+h.c.)
\end{equation}
Here, we denote the hopping integral between orbital $\alpha$ of unit cell $j$ and orbital $\beta$ of unit cell $l$ as $H_{jl}^{(\alpha\beta)}$, while its diagonal terms represent the site-energy of each orbital. A key characteristic of the low-energy ($E_F\pm k_BT$) electronic structure that determines the $T_c$ in the Migdal-Eliashberg theory is the Fermi-surface DOS $\rho_F$. Here, our 25-orbital model predicts $\rho_F = 0.932$, 0.894, and 0.881\,eV$^{-1}$ for 150\,GPa, 200\,GPa, and 250\,GPa LaH$_{10}$, respectively, aligning with findings by Errea \textit{et al.}\,\cite{errea2020quantum}.
 
Since the centers of charges are spatially separated among the Wannier orbitals, the light-matter interaction in the second-quantized expression can be described by the Peierls substitution $t_{jl}^{(\alpha\beta)} \rightarrow t_{jl}^{(\alpha\beta)}e^{i\mathbf{A}(t)\cdot(\mathbf{r}_j^{(\alpha)}-\mathbf{r}_l^{(\beta)})}$ for an instantaneous vector potential $\mathbf{A}(t)$. Here, we considered the steady state driven by a monochromatic laser, whose duration is much longer than its oscillation period. This steady-state condition can be mimicked by a periodic vector potential $\mathbf{A}(t) = \mathbf{A}_0\cos(\Omega t)$, where $A_0$ relates to the peak electric field strength by $E_0 = {\hbar\Omega A_0}/{ea_0}$. In this simplified situation, the dynamics governed by $\mathcal{H}(t)$ can be approximated using the Floquet theory, where the wavefunction solution forms a Fourier series\,\cite{jauho1994time,moskalets2002floquet,kohler2005driven}. Thus, the time-dependent Schrodinger equation can be mapped to an effective steady-state Hamiltonian $\mathcal{H}_{F}$ with extended dimensions in basis $\ket{m} = e^{im\Omega t}$ 
\begin{equation}\label{eq:HF}
    \begin{split}
        \mathcal{H}_F &= \sum_{m,n\in \mathbb{Z}}(\mathcal{H}^{(m-n)}-m\Omega\delta_{mn})\ket{m}\bra{n} \\
        &= \begin{pmatrix}
        \ddots & \ddots & \vdots & \vdots \\
        \ddots & \mathcal{H}^{(0)}+\Omega & \mathcal{H}^{(-1)} & \mathcal{H}^{(-2)} & \dots \\
        \dots & \mathcal{H}^{(1)} & \mathcal{H}^{(0)} & \mathcal{H}^{(-1)} & \dots \\
        \dots & \mathcal{H}^{(2)} & \mathcal{H}^{(1)} & \mathcal{H}^{(0)}-\Omega & \ddots \\
        & \vdots & \vdots & \ddots & \ddots
        \end{pmatrix}\,,
    \end{split}
\end{equation}
whose block matrices represent the Fourier transforms of $\mathcal{H}(t)$ [see Supplementary Note I]:
\begin{equation}\label{eq:Hm}
    \mathcal{H}^{(m)} = \sum_{jl}\sum_{\alpha\beta} i^m \mathcal{J}_m\left(\mathbf{A}_0\cdot(\mathbf{r}_j^{(\alpha)}\!-\!\mathbf{r}_l^{(\beta)})\right)H_{jl}^{(\alpha\beta)}(c_{l\beta}^\dagger c_{j\alpha}+h.c.)\,,
\end{equation}
with $\mathcal{J}_{m}(z)$ is the Bessel function of the first kind. In all pump conditions examined here, the 0th-order Bessel function dominates over all higher-order ones. Therefore, the DOS, and especially the Fermi-surface DOS $\rho_F$, are primarily determined by the diagonal blocks of Eq.~\eqref{eq:HF}, and we take this approximation in results below [see Supplementary Notes II and III for the discussion about higher-order Floquet analysis and justification of this approximation]. We analyze the non-thermal steady-state electronic structure and EPC using the Floquet Hamiltonian $\mathcal{H}_{F}$. 

We first consider the case at 220\,GPa, whose equilibrium properties have been studied by first-principles methods\,\cite{wang2019pressure}. The pump condition is chosen as the mid-IR laser ($\hbar\Omega = 400$\,meV), similar to that used in K$_3$C$_{60}$ experiments conducted in diamond anvil cell\,\cite{mitrano2016possible}. This photon energy corresponds to a 3\,$\mu$m laser, away from the absorption window of diamond, thereby ensuring that a relatively greater fluence can be delivered. Additionally, it is significantly higher than the $k_BT_c$ values examined in this paper, allowing the off-resonance approximations in the Floquet-DFPT calculations [see Supplementary Note II for detailed discussions]. Fig.~\ref{fig.band} shows the pump-amplitude dependence for the band structure, influenced by an $\hat{x}$-polarized laser. As the pump amplitude increases, the conduction bandwidth is gradually suppressed due to photon dressing. Consequently, the DOS $\rho_F$ increases from 0.898\,eV$^{-1}$ at equilibrium to 1.295\,eV$^{-1}$ for $E_0 = 80$\,mV/\AA\ (i.e. $\sim$ 4.5\,mJ/cm$^2$ for a 50-fs laser pulse). The overall increase of $\rho_F$ by 44\% signifies the potential of engineering LaH$_{10}$ electronic structure using the pump laser. Given that the $T_c$ from BCS theory is proportional to $\exp(-1/\rho_FV)$, with $V$ being the phonon-mediated interaction, the laser-enhanced DOS could foster a more robust superconducting state. 

\begin{figure}[!b]
    \centering
    \includegraphics[width=8.5cm]{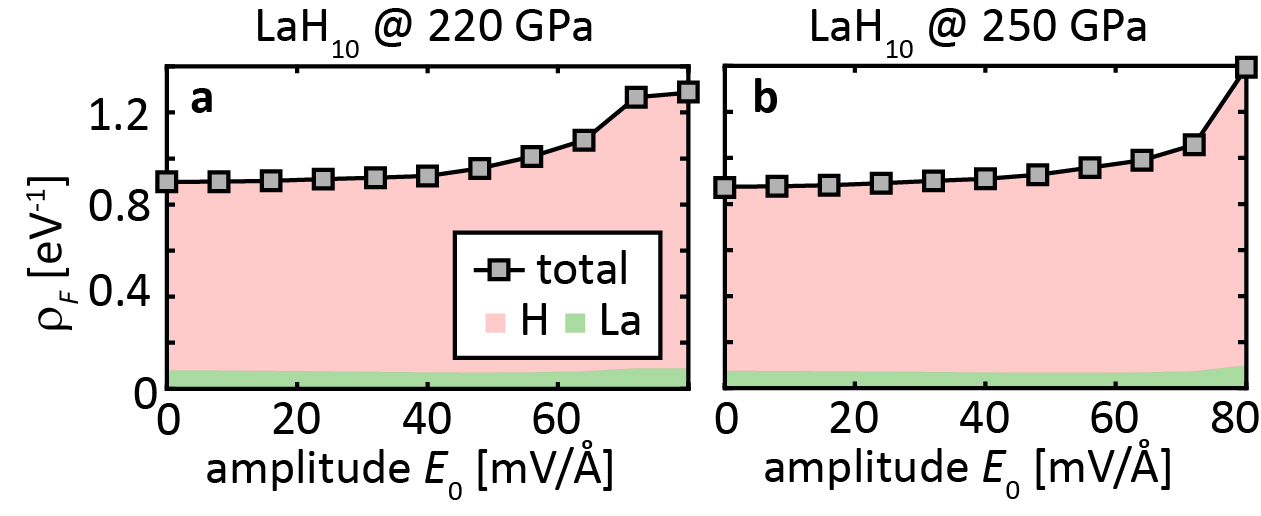	}\vspace{-3mm}
    \caption{\label{fig.dos}
        \textbf{Pump amplitude's influence on DOS and partial DOS contributions from H and La atoms.} \textbf{a} The DOS and partial DOS in LaH$_{10}$ at 220\,GPa under different pump amplitudes. The total DOS is depicted with the black lines and squares. Contributions from H orbitals are highlighted in red, while those from La orbitals are displayed in green. \textbf{b} Same as \textbf{a} but for LaH$_{10}$ at 250\,GPa.
    }    
\end{figure}

More specifically, the increase in $\rho_F$ occurs gradually for weaker pump amplitudes ($E_0 < 50$\,mV/\AA), rising by only 6\% up to 0.956\,eV$^{-1}$ [see Fig.~\ref{fig.dos}\textbf{a}]. This increase becomes more pronounced for pump amplitude $E_0$ exceeding 50\,mV/\AA. Such a reflection point in $\rho_F$ is primarily caused by the band flattening of the van Hove singularity (vHS) near the L point (indicated in the red dashed box of Fig.~\ref{fig.band}\textbf{a}). As depicted in Fig.~\ref{fig.band}, this vHS is 0.2\,eV below $E_F$ at equilibrium, and Floquet renormalization of the entire band shifts it across the Fermi level at $E_0 \sim 50$\,mV/\AA. The shift finishes at $E_0 \sim 70$\,mV/\AA, where the DOS plateaus. Additionally, two other hydrogen bands are renormalized from above $E_F$ to the Fermi level near the X point for pump amplitudes $E_0 > 70$\,mV/\AA. These bands may continue to increase $\rho_F$ but for unrealistically strong pumps beyond the range we are interested in. A similar result for $\rho_F$ is observed under 250\,GPa as well, except for a sharp increase around $E_0 = 70$\,mV/\AA [see Fig.~\ref{fig.dos}\textbf{b}]. This difference originates from the delayed upward shift of the band near the Fermi surface near the L point [see Supplementary Note IV for further discussions].

\begin{figure*}[!t]
    \centering
    \includegraphics[width=1\textwidth]{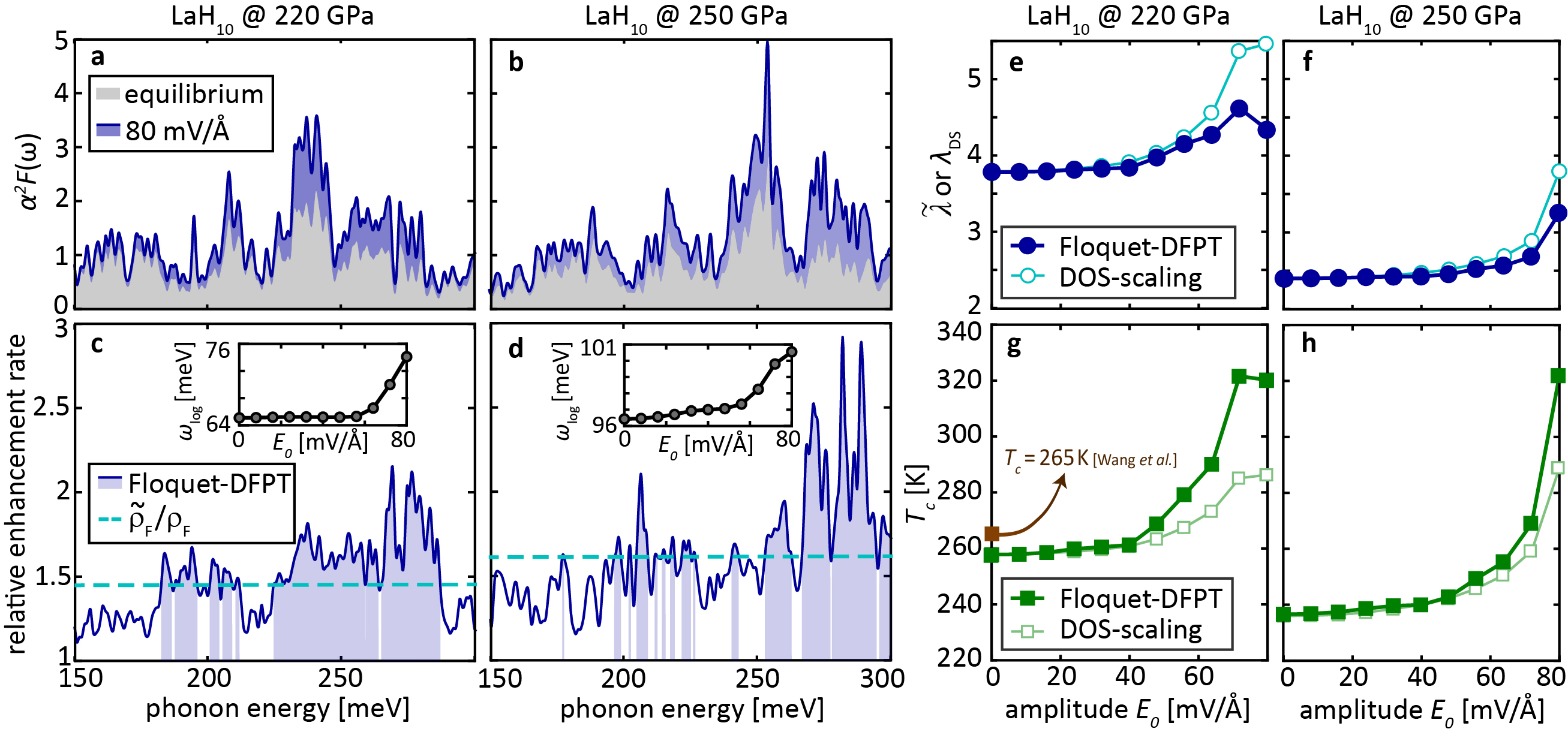	}\vspace{-3mm}
    \caption{\label{fig.lambda}
        \textbf{Floquet-engineered electron-phonon coupling and critical temperature.} \textbf{a} A comparison between equilibrium (gray) and Floquet-DFPT evaluated steady-state $\aF$ (blue) for LaH$_{10}$ at 220\,GPa. The steady-state example shown here is for an $\hat{x}$-polarized pump with $E_0 = 80$\,mV/\AA. \textbf{c} Phonon energy-dependent relative enhancement of $\aF$ in the $E_0 = 80$\,mV/\AA\, steady states at 220\,GPa compared against equilibrium $\aF$. The dashed line reflects the relative enhancement of $\aF(\omega)$ originating from the Floquet-engineered DOS ($\Tilde{\rho}_F(E_0)/\rho_F$) under the same pump condition, while the shaded regimes highlight the enhancement originating from increased EPC. The inset shows the weighted averaged phonon frequency $\omega_{\rm log}$ for varying pump amplitudes. \textbf{b}, \textbf{d} Similar to \textbf{a}, \textbf{c} but for LaH$_{10}$ at 250\,GPa. \textbf{e}-\textbf{f}: Pump amplitude dependence of $\lambda$ for the two pressures, respectively. Solid dots denote Floquet-DFPT simulation results (denoted as $\tilde{\lambda}$), while the open dots represent DOS-scaling approximations (denoted as $\lambda_{\mathrm{DS}}$). \textbf{g}-\textbf{h}: Pump amplitude dependence of $T_c$ for corresponding pressures, evaluated by Floquet-DFPT (solid squares) and DOS scaling (open squares). The equilibrium $T_c$ result from Ref.~\onlinecite{wang2019pressure} is highlighted in \textbf{g}. 
    }    
\end{figure*}

\subsection{Steady-State Electron-Phonon Coupling and Transition Temperature}\label{sec.b}

While the electronic structure analysis in Sec.~\ref{sec.a} motivates the possibility of increasing $T_c$, the latter is determined by multiple factors, especially the EPC. To accurately predict the transition temperature $T_c$ for the light-driven LaH$_{10}$, we further simulate the EPC matrix elements to obtain the Eliashberg spectral function $\aF$ in these nonequilibrium conditions. Our equilibrium phonon modes are obtained by DFPT implemented in \textsc{PHonon}, a module from \textsc{Quantum ESPRESSO} [see Supplementary Note V]. In the nonequilibrium laser-driven conditions, we assume that the phonon energies $\omega_{\mathbf{q}\nu}$ and eigen-modes remain unchanged in the non-thermal state, since the equilibrium states are obtained by structural optimization and stay far from a structural transition. The rationale behind this assumption is that the duration of pump pulses in experimental settings is typically in the order of 100\,fs. During such a short timescale, the energy transfer from the laser to the crystal lattice, which usually takes a few picoseconds, is unlikely to occur [see Supplementary Note VI for laser-induced corrections to phonons]. This assumption also prevents the decomposition effect of LaH$_{10}$ due to dehydrogenation over time\,\cite{liu2018dynamics,causse2023superionicity,wang2023quantum,zhou2025diffusion}.
That being said, the EPC's impact on superconductivity is through the steady-state coupling strength and the Eliashberg function.

To evaluate these quantities, we generalize the dimensionless EPC strength into the Floquet $\Tilde{\lambda}$, given by
\begin{equation}\label{eq:lambda}
    \Tilde{\lambda} = \int_0^\infty\frac{2}{\omega}\,\tildeaF(\omega)d\omega
\end{equation}
with steady-state Eliashberg spectral function defined as
\begin{equation}\label{eq:a2f} 
    \tildeaF(\omega) = \frac{1}{\Tilde{\rho}_F}\sum_{\mu\mu'\nu}\sum_{\mathbf{kk}'}\big|\Tilde{g}^{(\mu\mu')}_{\mathbf{kk}',\nu}\big|^2\delta(\Tilde{\varepsilon}_{\mu\mathbf{k}})\delta(\Tilde{\varepsilon}_{\mu'\mathbf{k}'})\delta(\omega-\omega_{\mathbf{q}\nu})\,.
\end{equation}
Here, $\mathbf{q} = \mathbf{k}-\mathbf{k}'$ and $\omega_{\mathbf{q}\nu}$ are phonon momentum vector and energy; $\Tilde{\varepsilon}_{\mu\mathbf{k}}$ is the Floquet band dispersion for band $\mu$; $\Tilde{\rho}_F$ is the Fermi-surface DOS for steady states. The simulation of the Floquet $\Tilde{\lambda}$ further requires evaluating the steady-state EPC matrix element 
\begin{equation}\label{eq:epcmatrix}
    \Tilde{g}^{(\mu\mu')}_{\mathbf{kk}',\nu} = \sqrt{\frac{\hbar}{2m_0\omega_{\mathbf{q}\nu}}}\braket{\Tilde{\psi}^{(\mu')}_{\mathbf{k}'}|\partial_{\mathbf{q}\nu}V|\Tilde{\psi}^{(\mu)}_{\mathbf{k}}}
\end{equation}
using Floquet-state-projected DFPT. As discussed in \textbf{Methods}, calculating Eq.~\eqref{eq:epcmatrix} is the most computationally intensive step, where the derivative of phonon self-consistent potential $\partial_{\mathbf{q}\nu}V$ (for wavevector $\mathbf{q}$ and branch $\nu$) is projected into a pair of Floquet eigen-states $\ket{\Tilde{\psi}^{(\mu)}_{\mathbf{k}}}$ (for band $\mu$ and momentum $\mathbf{k}$). We simulated $\Tilde{g}^{(\mu\mu')}_{\mathbf{kk}',\nu}$ and $\Tilde{\lambda}$ using a modified version of \textsc{EPW}\,\cite{lee2023electron}, an open-source module from \textsc{Quantum ESPRESSO} [See \textbf{Methods}].

In Fig.~\ref{fig.lambda}\textbf{a}, we compare the nonequilibrium $\tildeaF$ obtained for the maximum peak amplitude ($E_0 = 80$\,mV/\AA) with the equilibrium $\aF$ for 220\,GPa. Notably, the two $\delta$-functions for electronic dispersion scale with $\rho_F$. Therefore, the DOS increase of the Floquet-renormalized electronic structure leads to an augmented $\aF$ across almost all phonon energies. To depict this energy-dependent modification, we analyze the relative enhancement rate $\tildeaF/\aF$ for the same pump amplitude, as shown in Fig.~\ref{fig.lambda}\textbf{c}. This relative rate highlights the uneven enhancement of $\tildeaF$ for different phonon energies. Compared to the relative $\rho_F$ increase (dashed line), the simulated $\tildeaF$ is lower at some frequencies, indicating a net decrease in Fermi-surface-projected EPC matrix elements $\Tilde{g}^{(\mu\mu')}_{\mathbf{kk}',\nu}$ for these specific phonons. Conversely, in the 230-280\,meV phonon energy range, the Floquet-engineered electronic wavefunctions and changes in the Fermi surface shape lead to a pronounced increase in $\tildeaF(\omega)$, surpassing the overall scaling attributable to $\rho_F$. This increase reflects an enhanced effective EPC strength for phonons in the hydrogen optical branch, caused by the steady-state electronic structure. A direct consequence of this uneven enhancement is the notable increase in logarithmic-average frequency $\tilde{\omega}_{\log}$ displayed as the insets. Independent from the strength of $\lambda$, $\tilde{\omega}_{\log}$ plays a crucial role of the Debye frequency in the Allen-Dynes estimation of $T_c$\,\cite{allen1975transition}. Similar enhancement for EPC of high-energy phonons also occurs for the 250\,GPa states, whose $\aF$ predominantly increases at 260-300\,meV [see Fig.~\ref{fig.lambda}\textbf{b} and \textbf{d}]. 

The rise in $\aF$ naturally leads to a larger dimensionless EPC strength $\lambda$. As shown in Fig.~\ref{fig.lambda}\textbf{e}, the Floquet-state $\Tilde{\lambda}$ exhibits an accelerating growth with increasing pump amplitudes at 220\,GPa. The growth halts at $E_0 \sim 70$\,mV/\AA, where the vHS completes crossing the Fermi level. Compared to the 220\,GPa LaH$_{10}$, the $\Tilde{\lambda}$ for 250\,GPa is generally smaller for all pump amplitudes. This difference stems from their distinct structural instability at equilibrium. The 220\,GPa LaH$_{10}$ is closer to the structure phase transition at 210\,GPa, thus its phonons are softer compared to the 250\,GPa structure, leading to an overall increase in $\Tilde{\lambda}$ [see Figs.~\ref{fig.lambda}\textbf{a}-\textbf{b}]\,\cite{wang2019pressure}. A comprehensive discussion about light-engineered $\aF(\omega)$ and $\lambda$ is included in Supplementary Note V.

Within the BCS regime, the enhanced $\lambda$ typically indicates a higher $T_c$. Here, we estimate the nonequilibrium $T_c$ using the Allen-Dynes formula\,\cite{allen1975transition} with the Floquet steady-state $\Tilde{\lambda}$ and $\Tilde{\omega}_{\log}$ [see \textbf{Methods}]. Figs.~\ref{fig.lambda}\textbf{g}-\textbf{h} display the calculated $T_c$ for LaH$_{10}$ at 220\,GPa and 250\,GPa. Before adding the external laser field, the equilibrium LaH$_{10}$ yields a $T_c$ of 258\,K at 220\,GPa and 237\,K at 250\,GPa, consistent with previous studies\,\cite{wang2019pressure}. The increase in $\lambda$ raises $T_c$ from 258\,K at equilibrium to 320\,K for $E_0 = 80$\,mV/\AA\ at 220\,GPa, yielding a 25\% \rvs{estimated} increase. Similarly, $T_c$ rises from 237\,K to 322\,K yielding a 35\% \rvs{estimated}  increase for 250\,GPa. Both scenarios \rvs{could potentially} reach a room temperature $T_c$ (300\,K) under relatively strong pump conditions. (To clarify, ``strong pump condition'' refers to the relative strength of $E_0$ in Fig.~\ref{fig.lambda}. This does not compromise the validity of the approximations used in the Floquet-DFPT method.) The $T_c$ rise roughly parallels the increase in $\lambda$ shown in Figs.~\ref{fig.lambda}\textbf{c}-\textbf{d}. Notably, the $T_c$ for 220\,GPa LaH$_{10}$ plateaus, not decreasing in the strong pump regime ($E_0 > 70$\,mV/\AA), where $\lambda$ starts to drop. This is due to the aforementioned uneven enhancement of $\aF$, leading to a rise in $\Tilde{\omega}_{\log}$ and further contributing to the $T_c$ increase. 

To unravel various factors contributing to $T_c$ and understand the mechanism of light-enhanced SC, we further examine a simpler approximation for the Floquet renormalization of $\aF$ and $T_c$. Assuming that the EPC matrix elements remain constant compared to the equilibrium $g_{\mathbf{kk}',\nu}^{(\mu\mu')}$ and the Fermi surface is unchanged, the bandwidth renormalization described in Sec.~\ref{sec.a} would manifest as a uniform, phonon energy-independent rise in $\aF(\omega)$ (see dashed lines in Figs.~\ref{fig.lambda}\textbf{c}-\textbf{d}). That being said, this DOS scaling approximates steady-state quantities as
\begin{equation}\label{eq:scaling}
    \aF_{\rm DS}(\omega) = \frac{\Tilde{\rho}_F(E_0)}{\rho_F}\aF(\omega)\,,\, \Tilde{\lambda}_{\rm DS}(E_0) = \frac{\Tilde{\rho}_F(E_0)\lambda}{\rho_F}\,
\end{equation}
where $\aF$, $\lambda$ and $\rho_F$ are the corresponding quantities in equilibrium. Incorporating Eq.~\eqref{eq:scaling} into the Allen-Dynes formula [see Eq.~\ref{eq:ad} in \textbf{Methods}], the assessed $T_c$ then reflects solely the impact of Floquet-induced DOS increase, excluding nonequilibrium modifications to the EPC matrix elements. The open dots in Figs.~\ref{fig.lambda}\textbf{e}-\textbf{h} show this DOS-scaling simplification and are contrasted by the simulated $\Tilde{\lambda}$ shown in solid dots. This simplification overestimates $\lambda$, since it ignores the decrease in EPC for most phonon frequencies as discussed above (also see Supplementary Note V). The overestimation becomes more pronounced near the vHS (e.g. for $E_0 \sim 70$\,mV/\AA\ at 220\,GPa), where the Fermi-surface topology undergoes a transition. Interestingly, despite the up to 20\% overestimation of $\lambda$, the $T_c$ evaluated via this simplified scaling is always lower than that obtained from the Floquet-DFPT simulation (solid squares in Figs.~\ref{fig.lambda}\textbf{g}-\textbf{h}). This difference arises from the uneven increase of $\tildeaF$ across energy: the enhanced EPC coupling to high-energy (230-280\,meV) phonons elevates the $\tilde{\omega}_{\log}$, influencing $T_c$ in the Allen-Dynes formula apart from $\lambda$; this Floquet engineering of EPC is overlooked in the simplified scaling assessment, which evaluates $\tildeaF$ evenly in energy (see dashed lines in Figs.~\ref{fig.lambda}\textbf{c}-\textbf{d}).

\begin{figure}[!b]
    \centering
    \includegraphics[width=8.7cm]{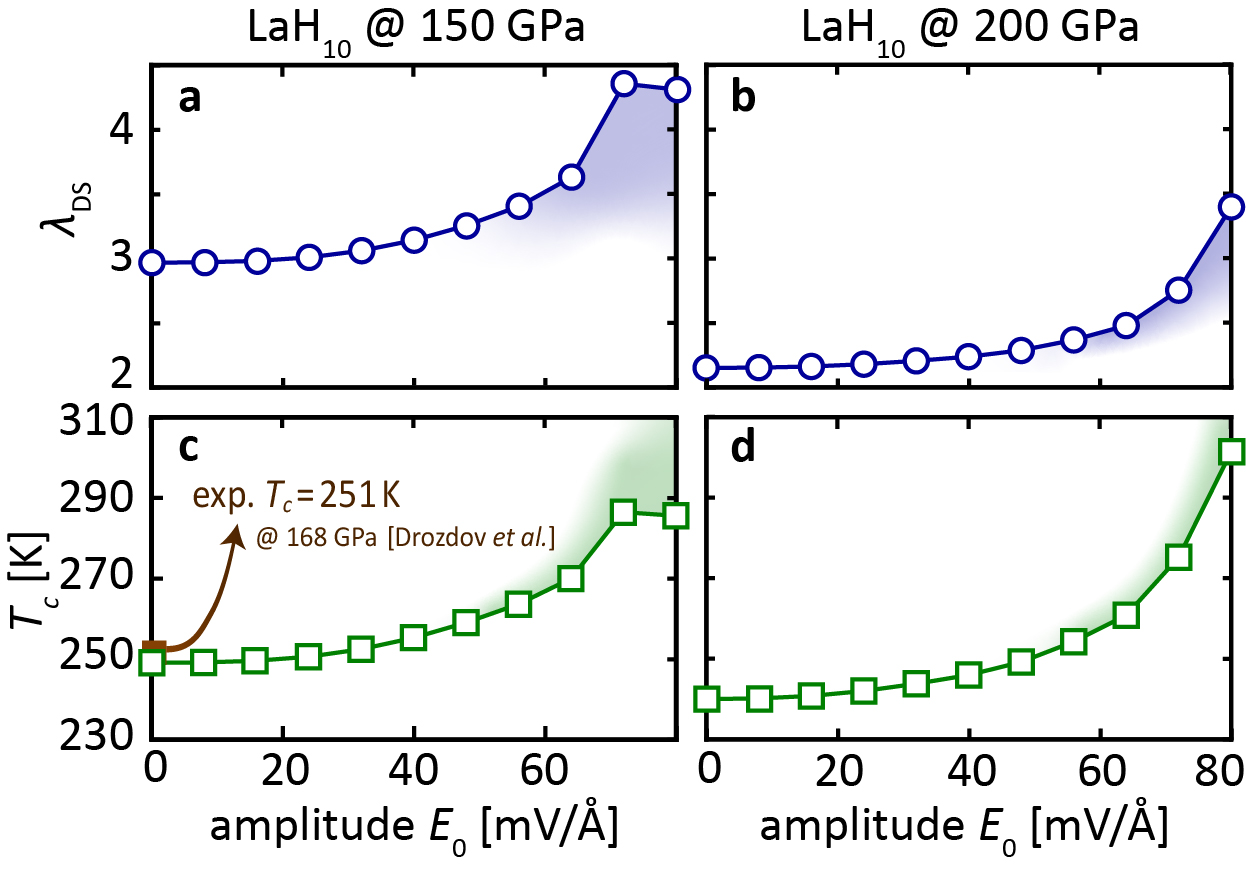	}\vspace{-3mm}
    \caption{
        \textbf{DOS-scaling approximated results for low pressures.} \textbf{a},\textbf{c} The approximated electron-phonon coupling $\Tilde{\lambda}_{\rm DS}$ and critical temperature $T_c$ under 150\,GPa, using the DOS-scaling method. The shaded regions indicate potential overestimations of $\lambda$ and underestimation\rvs{s} of $T_c$. The experimentally measured $T_c$ at a comparable pressure (168\,GPa), as reported in Ref.~\onlinecite{drozdov2019superconductivity}, is highlighted in \textbf{c}. \textbf{b},\textbf{d} Similar to \textbf{a} and \textbf{c} but for LaH$_{10}$ under 200\,GPa. 
    }
    \label{fig.press}
\end{figure}

\subsection{Light-Enhanced $T_c$ at Experimental Pressures}\label{sec.c}

While the above simulations suggest a 25\% to 35\% \rvs{potential} enhancement of $T_c$ in the nonequilibrium states of LaH$_{10}$ and indicate the feasibility of achieving room-temperature SC under realistic pump conditions, they are based on Floquet-DFPT simulations above 210\,GPa. Due to significant lattice anharmonicity below 210\,GPa, direct simulations become unreliable below this pressure\,\cite{errea2020quantum}, presenting additional challenges for our nonequilibrium simulations. However, experimentally accessible pressures typically fall below 200\,GPa\,\cite{drozdov2019superconductivity,somayazulu2019evidence}. To offer practical predictions for guiding the design of light-enhanced $T_c$ in experimentally relevant samples, we employ the previously mentioned DOS-scaling simplification as a rough estimate for the lower bound of the Floquet steady-state $T_c$. 

Specifically, we employed the equilibrium anharmonic phonon results from the well-recognized study by Errea \textit{et al.} for lower-pressure LaH$_{10}$\,\cite{errea2020quantum}. Without re-evaluating the configurational energy surface and EPC matrix elements for Floquet electronic states, we estimate the Floquet-engineered $\lambda$ and $T_c$ using Eqs.~\eqref{eq:scaling}. Here, we choose 150\,GPa and 200\,GPa LaH$_{10}$ as examples, whose experimental $T_c$s align closely with simulations incorporating anharmonicity\,\cite{drozdov2019superconductivity,somayazulu2019evidence}. By applying the same parameters and the Allen-Dynes formula, we reproduce experimentally measured $T_c$ reported in Ref.~\onlinecite{errea2020quantum}. 

The simulated dimensionless EPC $\lambda_{\rm DS}$ is represented by the blue circles in Figs.~\ref{fig.press}\textbf{a}-\textbf{b}. Analogous to the high-pressure scenarios discussed in Sec.~\ref{sec.b}, $\lambda_{\rm DS}$ generally increases with the Floquet renormalization of the electronic bandwidth and the enhancement of $\rho_F$. The relative increase for the most intense pump condition is between 45-60\% for both 150\,GPa and 200\,GPa, similar to the high-pressure results. Furthermore, Figs.~\ref{fig.press}\textbf{c}-\textbf{d} display the DOS-scaling estimated $T_c$'s for varying pump amplitudes of LaH$_{10}$ at 150\,GPa and 200\,GPa. As the pump amplitude increases from 0 to 80\,mV/\AA, the $T_c$ rises from 249\,K to 285\,K for 150\,GPa LaH$_{10}$ and from 240\,K to 301\,K for 200\,GPa LaH$_{10}$, approaching room temperature at the highest pump amplitude. It is important to note that the $T_c$ predicted using the DOS-scaling approach should not be regarded as an accurate prediction. Rather, as demonstrated in Fig.~\ref{fig.lambda}, it acts as the lower bound for the Floquet-DFPT simulated $T_c$, since the DOS scaling does not account for the increase in $\Tilde{\omega}_{\log}$. To give the best of our EPC-induced superconductive prediction, we added a guide-to-the-eye $\lambda$ and $T_c$ estimation, which may prove useful for further experimental investigations. 

It is important to note that the DOS-induced increase in $\lambda$ and $T_c$ is not unique to LaH$_{10}$ but also exists in several other hydrides as well. To demonstrate this, we examined the universality and feasibility of this behavior in other hydride superconductors, including YH$_{10}$, CeH$_9$ and CaH$_6$. In all cases, the band structures undergo compression, leading to a significant rise in $\lambda$ and $T_c$ driven by an increase in $\rho_F$. However, not all of these hydrides achieve superconductivity above room temperature. The detailed results are provided in Supplementary Note VII.

\subsection{Polarization Dependence}\label{sec.d}

\begin{figure}
    \centering
    \includegraphics[width=8.5cm]{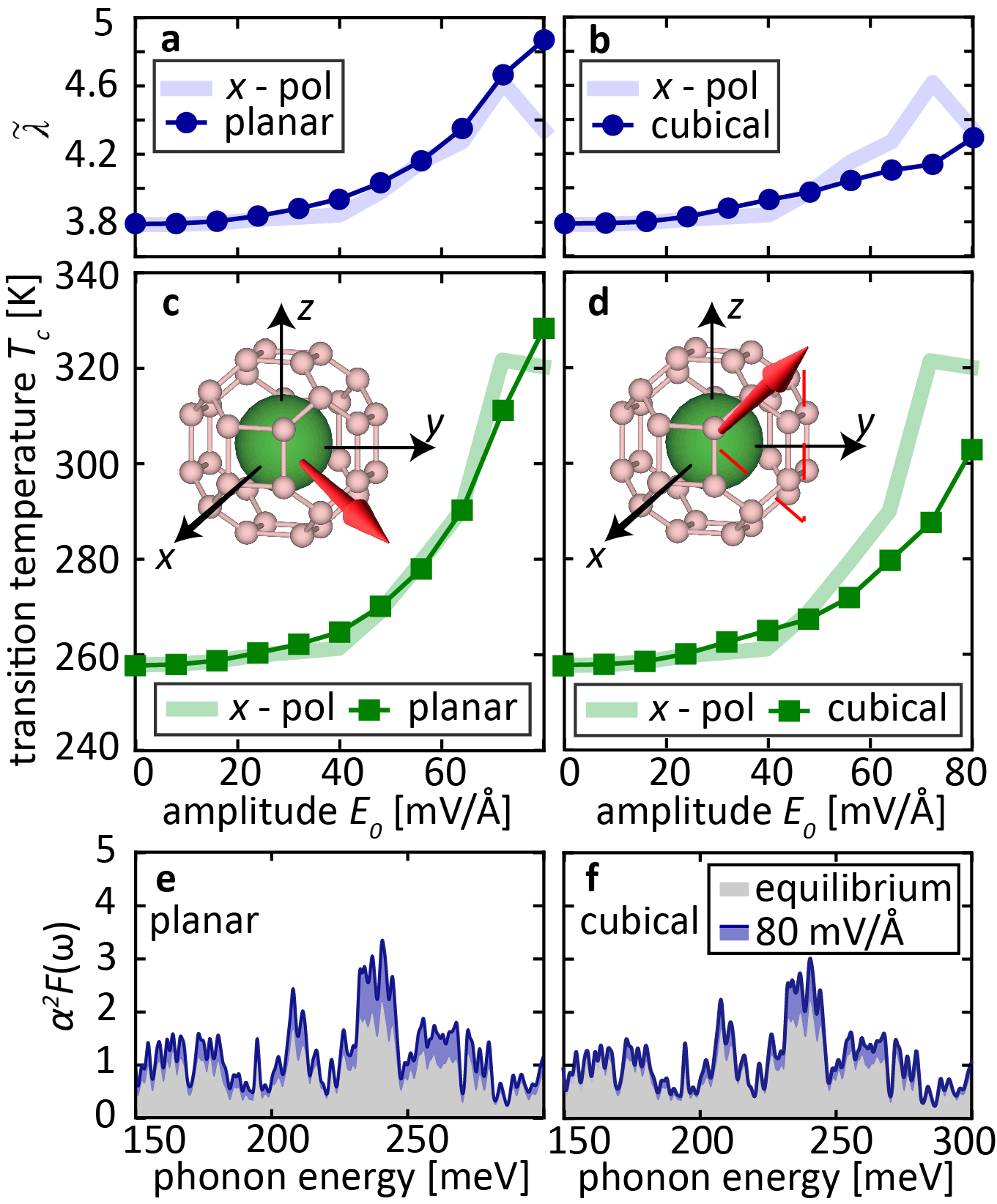	}\vspace{-2mm}
    \caption{
        \textbf{Polarized dependence of steady-state properties.} \textbf{a} The steady-state $\Tilde{\lambda}$ (solid circles) engineered by planar-diagonal-polarized pump with various pump amplitudes, simulated using Floquet-DFPT for LaH$_{10}$ under 220\,GPa. \textbf{b} Similar to \textbf{a} but for the cubical diagonal polarization. \textbf{c},\textbf{d} The simulated transition temperature $T_c$ corresponds to the pump conditions for \textbf{a} and \textbf{b}, respectively. The insets guide the eye for the two polarization directions. The light blue and green ribbons in \textbf{a}-\textbf{d} display comparative results obtained from x-polarized pump conditions, extracted from Figs.~\ref{fig.lambda}\textbf{e}-\textbf{g} (solid dots). \textbf{e},\textbf{f} A comparison between equilibrium (gray) and steady-state $\aF$ (blue) for LaH$_{10}$ under planar and cubical diagonal pump polarization, respectively, for $E_0 = 80$\,mV/\AA. 
    }
    \label{fig.yz}
\end{figure}

In the previous subsections, our exploration focused on the light-engineered band structure, EPC, and $T_c$ for LaH$_{10}$ specifically using an $\hat{x}$-polarized laser. In this section, we shift our focus to discuss the polarization dependence. Given the clathrate-like crystal structure of LaH$_{10}$, the polarization dependence should be minimal within a unit cell and predominantly influenced by the crystal symmetry \textit{Fm$\bar{3}$m}. Therefore, we examine the nonequivalent polarization planar diagonal (i.e., $\mathbf{A}_0={A_0}(\hat{x}+\hat{y})/\sqrt{2}$) and cubical diagonal (i.e., $\mathbf{A}_0={A_0}(\hat{x}+\hat{y}+\hat{z})/\sqrt{3}$). Continuing with the numerically stable examples at 220\,GPa and 250\,GPa, Figs.~\ref{fig.yz}\textbf{a}-\textbf{b} present the pump-amplitude dependence of $\Tilde{\lambda}$ for these two polarizations. Although the polarization dependence is relatively minor compared to the significant increase in $\Tilde{\lambda}$ by $\hat{x}$-polarized light, we observe that the $\Tilde{\lambda}$ induced by the planar-polarized laser is almost identical to that by the $\hat{x}$-polarized laser, as seen in Fig.~\ref{fig.lambda}. In contrast, the increase in $\Tilde{\lambda}$ is slower for the cubical-polarized laser. This distinction is further shown in the simulated $T_c$ in Figs.~\ref{fig.yz}\textbf{c}-\textbf{d}. The steady-state $T_c$ is 9\% lower in the cubical-polarized pump condition, compared to the $\hat{x}$- and planar-polarized pump conditions. If we only consider the relative enhancement compared to equilibrium, the cubical-polarized pump induces 30\% less enhancement compared to the other two polarizations.

To gain insight into the polarization-dependent behavior, we analyze the electronic structure across the entire Fermi surface. Since the contribution to DOS at each point on the Fermi surface is inversely related to the Fermi velocity at that point, we investigate the normal direction of the Fermi surface for the lowest $v_F$ (i.e.~the vHS), which aligns with the L-U cut and its 23 symmetric equivalents [see Supplementary Note VIII]. We find that the summed projection of three investigated polarizations ($\hat{x}$, planar diagonal and cubical diagonal) onto these vHS normal vectors is almost identical to each other. As shown in Supplementary Fig.~5, this homogeneity leads to the similarity in the steady-state $\Tilde{\lambda}$ and $T_c$ evolution among these three different polarizations. 

Consequently, the polarization dependence reported in Figs.~\ref{fig.yz}\textbf{a}-\textbf{d} can be attributed to the Floquet modification of EPC. We then analyze the Eliashberg spectral functions under these various pump polarizations. Fig.~\ref{fig.yz}\textbf{e} shows that the light-induced changes in $\aF$ for $E_0 = 80$\,mV/\AA\ are nearly identical to those from the $\hat{x}$-polarized pump [see Fig.~\ref{fig.lambda}\textbf{a}], correlating with their similar $T_c$ evolutions presented in Fig.~\ref{fig.yz}\textbf{c}. In contrast, for the cubical-diagonal-polarized pump, the increase in $\aF$ under all frequencies is less pronounced compared to the other two polarizations. This difference is particularly noticeable in the range of 230-270\,meV, which is expected to significantly influence the $T_c$ as discussed in Sec.~\ref{sec.b}. This observation aligns with the less efficient light-induced superconductivity observed under the cubical-polarized pump condition, as shown in Fig.~\ref{fig.yz}\textbf{d}.

\section{Discussion}\label{discussion}

In our study, we have explored a dynamical approach to potentially increase the superconducting transition temperature ($T_c$) in hydride materials, exemplified by LaH$_{10}$. In the Floquet state induced by light, both the augmented DOS and the upward shift in average phonon energy significantly contribute to elevating $T_c$ by 20-30\%, potentially raising it above room temperature. For lower-pressure compounds, our predictions offer a lower-bound estimate for $T_c$, suggesting the possibility for even more effective light-enhanced superconductivity. More generally, our Floquet-DFPT simulation paves the way for understanding dynamical superconductivity in light-element materials and the nonequilibrium electron-phonon dynamics driven by laser light.

It is important to recognize that the Floquet theory we have employed is an approximation of light-driven nonequilibrium dynamics. There have been several successful demonstrations in various materials\,\cite{wang2013observation,mahmood2016selective,shan2021giant,ito2023build,oka2009photovoltaic,mciver2020light,zhou2023pseudospin}. However, achieving non-thermal Floquet states in experiments remains a significant challenge\,\cite{delatorre2021colloquium}. In our assumptions based on Floquet theory, we consider the pump to be off-resonant with the relevant electronic states and sufficiently weak, allowing us to approximate the Floquet states at their lowest order. Although it is theoretically possible to evaluate full Floquet states including all higher-order terms [see Supplementary Note II], the Allen-Dynes-based simulation for $T_c$ necessitates a specific electronic state distribution resembling a Fermi sea, which cannot be accurately determined at infinite orders. For the parameters used in this work, the Fermi-Dirac distribution in Floquet steady states shows a majority of over 70\% similarity with electronic eigenstates near Fermi level derived from real-time dynamics, with the minimum value never dropping below 65\%, thereby validating our assumption regarding electronic states near Fermi surface.

In our simulation of hydride superconductivity, like many current studies, we face the challenge of accurately depicting multi-reference effects, a known difficulty within the Kohn-Sham DFT framework\,\cite{huang2016much}. The influence of Coulomb interaction on electron pairing is only addressed perturbatively through the Morel-Anderson pseudopotential $\mu^*$. Moreover, the Migdal-Eliashberg theory starts to deviate from a valid description of superconductivity when $T_c$ reaches $0.1 \omega_{\rm Debye}$\,\cite{esterlis2018a}, which for hydrides is approximately $330$\,K.  Given this boundary of validity, along with benchmarks between first-principles methods, Migdal-Eliashberg theory, and experimental results\,\cite{drozdov2019superconductivity,somayazulu2019evidence,sun2021high}, our perturbative approximations are justifiable within the current context. However, it is important to note that quantum fluctuations become increasingly significant when $T_c$ approaches or surpasses $330$\,K\,\cite{esterlis2018breakdown}. Extending this method to superconducting materials with either strong correlations or low Debye frequencies also requires a more precise correction for quantum many-body effects.

\section{Methods}

\subsection{First-Principles Calculations for Electronic Structure}

We employ the \textsc{PWscf} module in the \textsc{Quantum ESPRESSO} package\,\cite{giannozzi2017advanced} to simulate the electronic structure of LaH$_{10}$ under hydrostatic pressure, based on the plane-wave pseudopotential method. We use the Perdew-Burke-Ernzerhof (PBE) exchange-correlation functional\,\cite{perdew1997generalized} with SG-15 optimized norm-conserving Vanderbilt (ONCV) pseudopotentials\,\cite{hamann2013optimized}. To expand the wavefunctions, a cutoff energy of 80 Rydberg is used and integrate over the Brillouin zone via a $\Gamma$-centered Monkhorst-Pack 24$\times$24$\times$24 grid\,\cite{monkhorst1976special}. The convergence criteria of ionic structure relaxation under pressure and the electronic self-consistent field are set to $10^{-4}$ Rydberg/Bohr and $10^{-8}$ Rydberg, respectively. 

The Wannier orbitals are obtained by an additional non-self-consistent calculation with a 6$\times$6$\times$6 $\mathbf{k}$-grid. The simulated eigen wavefunctions are used to construct maximally localized Wannier functions (MLWFs) through the \textsc{Wannier90} package\,\cite{pizzi2020wannier90}. This Wannier downfolding procedure takes 10 H $s$ orbitals, 3 La $p$ orbitals, 5 La $d$ orbitals and 7 La $f$ orbitals as the initial projections, resulting in a tight-binding model whose band structure reproduces the DFT results. A total number of 47 bands are involved during the disentanglement procedure. The DOS is obtained by integrating the band distribution in a 72$\times$72$\times$72 $\mathbf{k}$-grid, with Gaussian broadening of 0.1\,eV. 

\subsection{Electron-Phonon Coupling Calculations}

We employ DFPT implemented in \textsc{PHonon} to calculate the phonon dispersion of \textit{Fm$\bar{3}$m} LaH$_{10}$ under different pressure conditions. A $\Gamma$-centered Monkhorst-Pack sampling\,\cite{monkhorst1976special} of 6$\times$6$\times$6 is adopted as a uniform grid of $\mathbf{q}$-vectors and the convergence criteria of phonon calculation are set to $10^{-16}$ Rydberg. The EPC matrix elements are simulated using \textsc{EPW}\,\cite{lee2023electron} in equilibrium. The steady-state simulations are realized by substituting the Floquet steady-state wavefunctions, obtained by diagonalizing the Floquet-Wannier Hamiltonian $\mathcal{H}_F$, into the finer grid rotation matrices $U_{\mathbf{k}}$ for EPC matrix elements $\Tilde{g}_{\mathbf{k}\mathbf{k}',\nu}^{(\mu\mu')} = \sqrt{\frac{\hbar}{2m_0\omega_{\mathbf{q}\nu}}}\bra{\Tilde{\psi}^{(\mu')}_{\mathbf{k}'}}\partial_{\mathbf{q}\nu}V\ket{\Tilde{\psi}^{(\mu)}_{\mathbf{k}}}$. The phonon momentum $\mathbf{q}$ is sampled over a uniform Fourier-interpolated 12$\times$12$\times$12 grid. 

\subsection{Superconductivity Transition Temperature}

We calculated the estimated corresponding nonequilibrium results with the full Allen-Dynes formula\,\cite{allen1975transition}. Using this formula, the transition temperature $T_c$ is given by
\begin{equation}\label{eq:ad}
    T_c = \omega_{\log}\frac{f_1f_2}{1.2}\exp\left({\frac{-1.04(1+\lambda)}{\lambda-\mu^*(1+0.62\lambda)}}\right)
\end{equation}
with
\begin{eqnarray}
    f_1f_2 = && \sqrt[3]{1+\left[\frac{\lambda}{2.46(1+3.8\mu^*)}\right]^{3/2}}\nonumber \\
    && \cdot\left[1-\frac{\lambda^2(1-\omega_2/\omega_{\log})}{\lambda^2+3.312(1+6.3\mu^*)^2}\right]
\end{eqnarray}
The logarithmic average frequency $\omega_{\log}$ and mean-square frequency $\omega_2$ are calculated as
\begin{equation}
    \omega_{\log} = \exp\left[\frac{2}{\lambda}\int_0^{\infty}\frac{\aF(\omega)}{\omega}\log{\omega}\,d\omega\right]
\end{equation}
\begin{equation}
    \omega_2 = \sqrt{\frac{1}{\lambda}\int_0^{\infty}\frac{2\aF(\omega)}{\omega}\omega^2d\omega}
\end{equation}
Following Ref.~\onlinecite{wang2019pressure}, we took the typical Morel-Anderson Coulomb pseudopotential $\mu^* = 0.1$ throughout this paper.

\section*{Acknowledgements}

The authors thank Cheng-Chien Chen, Ilya Esterlis, Yu He, Matteo Mitrano, and Michael Sentef for insightful discussions. This work is supported by the Air Force Office of Scientific Research Young Investigator Program under grant FA9550-23-1-0153. W.-C.C. and Y.W. also acknowledge support from the National Science Foundation (NSF) award DMR-2132338. A.D.S acknowledges support from the NSF awards No.~OIA-2148653 and DMR-2142801. Simulation results were obtained using the Frontera computing system at the Texas Advanced Computing Center. Frontera is made possible by NSF Award No.~OAC-1818253.

\section*{Data Availability}

The data supporting the findings of this study are available in the public repository Figshare at \href{https://doi.org/10.6084/m9.figshare.29500673}{10.6084/m9.figshare.29500673}.
%All the data generated in this work are available upon reasonable request from C.X., A.D.S. and Y.W.

\section*{Code Availability}

\textsc{Quantum ESPRESSO} is an open-source suite of computational tools available at \href{quantum-espresso.org}{quantum-espresso.org}, with an interface to open-source packages like \textsc{Wannier90} (\href{wannier.org}{wannier.org}) and \textsc{EPW} (\href{epw-code.org}{epw-code.org}). 

\section*{Author Contributions} 

Y.W. conceived the project. C.X. and A.D.S. designed the code. C.X. and H.Y. performed the calculations and analyzed the data, with the help of W.-C.C. Y.W., C.X., A.D.S., and H.Y. wrote the manuscript.

\section*{Competing Interests} 

The authors declare no competing interests.

\bibliography{references}
\end{document}